\documentclass[twocolumn]{aastex63}
\usepackage{color, soul}

\begin{document}

\title{Radio Counterpart Candidates to 4FGL-DR2 Unassociated Sources}

\author[0000-0001-7887-1912]{S. Bruzewski}
\affiliation{Department of Physics and Astronomy, University of New Mexico, Albuquerque, NM 87131, USA}

\author[0000-0001-6672-128X]{F.K. Schinzel}
\altaffiliation{An Adjunct Professor at the University of New Mexico.}
\affiliation{National Radio Astronomy Observatory, P.O. Box O, Socorro, NM 87801, USA}

\author[0000-0001-6495-7731]{G.B. Taylor}
\affiliation{Department of Physics and Astronomy, University of New Mexico, Albuquerque, NM 87131, USA}

\author[0000-0001-9737-9667]{L. Petrov}
\affiliation{NASA, Goddard Space Flight Center, 8800 Greenbelt Rd, Greenbelt MD 20771, USA}

\correspondingauthor{S. Bruzewski}
\email{bruzewskis@unm.edu}

(\received{2020 02 14})

\begin{abstract}
    For the duration of the \textit{Fermi Gamma-Ray Space Telescope}'s mission, approximately one-third of the point sources detected have been noted as "unassociated," meaning that they seem to have no known counterpart at any other wavelength/frequency. This mysterious part of the gamma-ray sky is perhaps one of the largest unknowns in current astronomical pursuits, and as such has been probed extensively by various techniques at various frequencies. Radio frequencies have been perhaps one of the most fruitful, producing a large fraction of the identified and associated Active Galactic Nuclei (AGN) and pulsars noted in each update of the point source catalogs. Here we present a catalog of 7432 radio counterpart candidates for unassociated gamma-ray fields in the 2nd Data Release of the 4th Fermi Point Source Catalog (4FGL-DR2). A description of the catalog and source types is provided followed by a discussion that demonstrates how the results of this work will aid new associations and identifications. As part of this work, we also present a first catalog derived from "quicklook" images of the Very Large Array Sky Survey (VLASS).
\end{abstract}


\section{Introduction}

In the time since its launch in 2008, the \textit{Fermi Gamma-Ray Space Telescope} has revolutionized the $\gamma$-ray regime of astrophysics, increasing the number of such high energy sources by more than an order of magnitude. However, it is important to note that since the first data release, approximately 30\% of its sources have been classified as "unassociated," meaning there exists no known counterpart in any other electromagnetic regime. In the Large Area Telescope 10-year Source Catalog \citep[4FGL-DR2, see][]{2020arXiv200511208B}, the most recent data release at the time of writing, 1705 of the total 5788 sources (29\%) are labeled as such. Currently, the $\gamma$-ray sky may be the least understood sector in all of electromagnetic astronomy. 

This largely mysterious collection of sources has served as a tantalizing target for searches at various energy regimes. Thus far it could perhaps be said that the most success has been had at radio frequencies, where a strong physical connection exists between the radio and the $\gamma$-ray sky through non-thermal emission processes. Furthermore, source densities on the sky are relatively comparable. This is because associations are most typically found by (1) matching variability, or (2) sky proximity. In the latter case, one must be able to show that the likelihood of finding a radio source with a particular feature set, when compared to the likelihood of finding a random background source, is sufficiently high as to warrant association. Details for this sort of association criteria are described more thoroughly in Section 5 of \citet{2020ApJS..247...33A} \citep[slightly updated in][]{2020arXiv200511208B}.

An ongoing effort by our group has been to provide associations via this method following each major \textit{Fermi} data release. In \citet{2013MNRAS.432.1294P} and \citet{2015ApJS..217....4S, 2017ApJ...838..139S}, observations are first performed with arrays and system configurations which allow for an entire \textit{Fermi} source's 95\% confidence positional uncertainty ellipse to fall inside a typical beam radius of a radio telescope. This allows for the rapid location of all radio sources in these fields which are sufficiently bright at arcsecond resolutions as to warrant followup at milliarcsecond resolution using Very Long Baseline Interferometry (VLBI). A bright nearby source detected via VLBI is much more likely to provide an association, as the number of chance background sources will be significantly lower, and milliarcsecond emission from a radio source associated with a $\gamma$-ray source is typically indicative of parsec-scale emission from an Active Galactic Nuclei (AGN). For example: approximately 85\% of $\gamma$-ray sources in the fourth catalog of AGN detected by \textit{Fermi} \citep[4LAC, see][]{2020ApJ...892..105A} have a VLBI counterpart, and for a majority of these, these associations were provided by the previous works described above.

In this paper we continue upon this previous work by providing a list of radio sources identified inside 4FGL-DR2 unassociated sources' 95\% confidence positional uncertainty ellipses (from here simply referred to as source ellipses), as well as meta information on the relative coverage of all unassociated sources in 4FGL-DR2 by this and all previous surveys as part of this effort. One other important aspect of this previous work was that the observations performed using connected element arrays (in this case the Karl G. Jansky Very Large Array (VLA) and the Australian Telescope Compact Array (ATCA)) were typically made with a split bandwidth setup. This method provides individual flux values at 5 and 7 GHz for sources, which in turn allows for very rudimentary spectral analysis of sources. We further improve upon this idea in this work by including data from the Very Large Array Sky Survey \citep{2020PASP..132c5001L}, which provides another data point for many sources at 3.0 GHz, expanding our spectral coverage dramatically, and allowing for more thorough spectral analysis.

In Section \ref{sec: observations} we describe the observations which were used to create the catalogs presented in this paper, whereas Section \ref{sec: methods} discusses the data reduction and novel analysis that generated said catalogs. In Section \ref{sec: results} we highlight individual groups of sources in the catalog with interesting features, and finally in Section \ref{sec: discussion} we conclude and discuss future prospects.

\section{Observations} \label{sec: observations}

This publication makes use of dedicated VLA observations, using the C-band receiver (4-8\, GHz) and data from the Very Large Array Sky Survey (VLASS), which are described in more detail in the following.

\begin{deluxetable}{lcccc}
\tablecaption{Summary of VLA Observations}
\tablehead{\colhead{Project} & \colhead{Start Date} & \colhead{Config} & \colhead{Hours} & \colhead{Pointings}}
\startdata
S5272   & 2012-10-26 & A & 9.0  & 271 \\
S7104   & 2015-03-16 & B & 9.0  & 384 \\
15A-466 & 2015-04-16 & B & 1.0  & 30 \\
18A-098 & 2018-04-13 & A & 26.0 & 1144 \\
SB072   & 2019-09-10 & A & 12.0 & 627 \\
\enddata
\tablecomments{\textit{Project}: For S-codes details can be found under approved \textit{Fermi} guest investigator proposals, regular VLA proposals can be looked up at http://library.nrao.edu/proposals; \textit{Start Date}: Date of the first observation under this project; \textit{Config}: VLA configuration code; \textit{Hours}: Total hours observed under project, including overheads; \textit{Pointings}: Number of individual target sky pointings per project.}
\end{deluxetable} \label{tbl: observations}

\subsection{Dedicated 5 and 7 GHz Observations}

This work combines our previous VLA observations targeting unassociated \textit{Fermi}/Large Area Telescope detected objects reported in the \textit{Fermi} Large Area Telescope second (2FGL) and third (3FGL) source catalogs with new dedicated observations of newly discovered unassociated $\gamma$-ray objects reported in the fourth source catalog (4FGL), including data release 2 (4FGL-DR2). Table~\ref{tbl: observations} summarized the key parameters of the conducted VLA observations with a total of 57 hours, including overheads, and 2456 pointings, corresponding to a covered sky area of approximately 43 square degrees at 6.0\,GHz. The observations between 2012 and 2015 were discussed in the context of 2FGL and 3FGL by \citep{2015ApJS..217....4S, 2017ApJ...838..139S}. 

In the case of S5272, S7104, 15A-466, and 18A-098, each pointing corresponded to a unique unassociated $\gamma$-ray source, with the pointing center defined by the $\gamma$-ray localization reported in the corresponding catalog. For 18A-098 a preliminary version of the 4FGL catalog was used. In some cases these pointings only covered the central part of the $\gamma$-ray localization error ellipse. In the case of SB072, we developed a new Python scheduling script which increased the overall efficiency of our survey and set up multi-pointing mosaics for sources where the primary beam did not fully encompass the localization ellipse. In this case all observations covered at least the entire 95\% localization error ellipse, up to those sources that would have required more than 7 pointings and which were excluded for time purposes. 

All observations used the VLA C-band receiver and a default continuum correlator setup using 3-bit digitizers, providing a maximum bandwidth of 4\,GHz split into two basebands of 2\,GHz bandwidth each, with center frequencies of 5.0 and 7.0\,GHz. Only in the case of S5272, 8-bit digitizers were used, which provided a maximum bandwidth of 2\,GHz, again split into two basebands with center frequencies of 5.0 and 7.3\,GHz. At the beginning of each observing segment one of the three primary flux density calibrators 3C\,48, 3C\,138, or 3C\,286 were observed with which flux density was scaled and instrumental bandpass were removed. In most cases, each target pointing was observed for about 30-45\,s. Nearby phase calibrators were added with typical integration times of about 15-30\,s each in order to solve for changes in the complex gains during the target observation. Observations were performed in VLA's A and B configuration, which provide antenna spacings ranging from 680\,m up to 36.4\,km in the case of A configuration and 210\,m to 11.1\,km in the case of B configuration. This corresponds to a synthesized beamwidth at half-power at 6.0\,GHz of about 0.33" and 1.0" for A and B configurations respectively.

The data calibration and analysis procedure is similar to that described in \citet{2015ApJS..217....4S,2017ApJ...838..139S}. The most recent data used a newer release of the Common Astronomy Software Applications (CASA). The data calibration and imaging was performed entirely within CASA. All imaging was performed using the same method as described in \citet{2017ApJ...838..139S}, generating an image for each baseband. In contrast to \citet{2015ApJS..217....4S,2017ApJ...838..139S}, source finding was performed using the Python Blob Detector and Source Finder (PyBDSF), more details can be found in Section \ref{sec: methods}.

\subsection{VLASS}

The Very Large Array Sky Survey \citep[VLASS, see][]{2020PASP..132c5001L} is an ongoing effort to survey the entire northern sky ($\delta>-40$) at 3.5 GHz, separated into three epochs. This is the first large scale survey performed by the instrument since its upgrades under the expanded Very Large Array project \citep{2011ApJ...739L...1P}, and as such will provide one of the most detailed views of the radio sky. At present, the most readily available data products from the project are so called "quicklook" images from first observation epoch, which are produced via an automated data reduction pipeline, and lack the full spectral and polarization information which will be included in the final data products. 

We provide, in the supplementary data alongside this paper, the first source catalog produced from these first epoch quicklook images. The exact details of how the catalog was produced, and what systematics were corrected for, can be found in Appendix \ref{apn: vlass}. This catalog was then folded into our analysis, alongside the source catalogs for the previously discussed 5 and 7 GHz observations.

\section{Methodology} \label{sec: methods}

The process for identifying sources in both VLASS quicklook data and our 5-7 GHz observations is largely the same. To begin with, source extraction is performed on all individual images using the Python implementation of PyBDSF \citep{2015ascl.soft02007M}. This is in contrast to the automated source extraction performed in previous iterations of this work, which used the Search And Destroy (SAD) task from the Astrophysical Image Processing System \citep[AIPS, see][]{1990apaa.conf..125G} to find a number of potential sources in an image and fit Gaussian components to these. We also note that data from previous projects was reprocessed via this new source extraction, so as to generate a uniform catalog of all sources identified in unassociated ellipses over time.

We then combine all of these individual image catalogs into a singular catalog for each frequency. These frequency catalogs then require further processing, as in both VLASS and our 5-7 GHz observations we expect some degree of overlap between images in certain cases, meaning we must seek and remove redundant catalog entries. In VLASS, further caution must be taken, as certain seemingly redundant entries may actually be from observations separated enough in time as to perhaps provide interesting data, whereas the 5-7 GHz overlaps likely happen in a very short time period of a single observation. In this case, or when the VLASS catalog entries are not significantly temporally separated, a simple criterion is used: if two sources in separate images are within a one arcsecond cutoff radius of each other, the entry from the image with worse RMS noise is given a flag noting it as the redundant entry of the pair.

One further consideration must be made for the dedicated observations: some sources in an individual observation may be located far from the pointing center, at distances which begin to approach the limit of the typical circularly-symmetric polynomial description of the beam. With this in mind, a Boolean flag is applied to any sources outside the 10\% beam power radius to mark their flux as lower limits in the final table (see 'IsLowerLim\_D5GHZ' and 'IsLowerLim\_D7GHZ' in Table \ref{tbl: f357cols}).

\begin{deluxetable*}{lccl}
\caption{F357 Columns}
\tablehead{\colhead{Name} & \colhead{Format} & \colhead{Unit} & \colhead{Description}}
\startdata
Name & 24s &  & Generated source name \\
RA & 3.4f & deg & Right Ascension of source (J2000) \\
RAErr & 1.2e & deg & Error in Right Ascension \\
Dec & 2.5f & deg & Declination of source (J2000) \\
DecErr & 1.2e & deg & Error in Declination \\
PeakFlux\_VLASS & 1.2e & mJy / beam & Peak flux density of VLASS (3.0 GHz) source \\
PeakFluxErr\_VLASS & 1.2e & mJy / beam & Error in peak flux density\\
IntFlux\_VLASS & 1.2e & mJy & Integrated flux density of VLASS (3.0 GHz) source \\
IntFluxErr\_VLASS & 1.2e & mJy & Error in integrated flux density\\
PeakFlux\_D5GHZ & 1.2e & mJy / beam & Peak flux density of D5GHZ (5.0 GHz) source \\
PeakFluxErr\_D5GHZ & 1.2e & mJy / beam & Error in peak flux density\\
IntFlux\_D5GHZ & 1.2e & mJy & Integrated flux density of D5GHZ (5.0 GHz) source \\
IntFluxErr\_D5GHZ & 1.2e & mJy & Error in integrated flux density\\
IsLowerLim\_D5GHZ &  &  & Given flux values may be treated as lower limits \\
PeakFlux\_D7GHZ & 1.2e & mJy / beam & Peak flux density of D7GHZ (7.0 GHz) source \\
PeakFluxErr\_D7GHZ & 1.2e & mJy / beam & Error in peak flux \\
IntFlux\_D7GHZ & 1.2e & mJy & Integrated flux density of D7GHZ (7.0 GHz) source \\
IntFluxErr\_D7GHZ & 1.2e & mJy & Error in integrated flux density\\
IsLowerLim\_D7GHZ &  &  & Given flux values may be treated as lower limits \\
Alpha & 2.5f &  & Calculated or fit spectral index of source \\
AlphaErr & 2.5f &  & Error in spectral index \\
FermiName & 17s &  & Name of encompassing Fermi source \\
MatchGAIA & 19s &  & GAIA crossmatch (1 arcsec cutoff) \\
DistGAIA & 2.5f & arcsec & Distance to GAIA source \\
MatchAllWISE & 19s &  & AllWISE crossmatch (1 arcsec cutoff) \\
DistAllWISE & 2.5f & arcsec & Distance to AllWISE source \\
Match2RXS & 21s &  & 2RXS crossmatch (11 arcsec cutoff) \\
Dist2RXS & 2.5f & arcsec & Distance to 2RXS source \\
Match2SXPS & 22s &  & 2SXPS crossmatch (13 arcsec cutoff) \\
Dist2SXPS & 2.5f & arcsec & Distance to 2SXPS source \\
MatchCSCR2 & 22s &  & CSCR2 crossmatch (9 arcsec cutoff) \\
DistCSCR2 & 2.5f & arcsec & Distance to CSCR2 source \\
IsResolved &  &  & Indicates resolved graph structure \\
TypeCode & 3s &  & Type codes (PyBDSF) for VLASS, D5GHZ, and D7GHZ \\
Class & 3s &  & Spectral classification of source
\enddata
\tablecomments{Here the format column refers specifically to the Python string formatting code used to generate the final output table. Boolean columns have no specified formatting.}
\end{deluxetable*}
 \label{tbl: f357cols}

To combine these three catalogs into our final data product, we iterate through each unassociated Fermi source, identifying all sources from VLASS, our Dedicated 5 GHz (D5GHZ), and our Dedicated 7 GHZ (D7GHZ) observations which lay inside the 95\% confidence positional uncertainty ellipse. We then begin grouping sources.

Sources in the same catalog are linked if they have the same Island ID generated by PyBDSF, allowing us to track structure denoted by the source extractor. Sources in different catalogs are linked using a unitless uncertainty-scaled distance metric 

\begin{equation}
    r = \sqrt{ \frac{(\alpha_1-\alpha_1)^2 \cos^2{((\delta_1+\delta_2)/2)}}{\sigma_{\alpha,1}^2 + \sigma_{\alpha,2}^2}  + \frac{(\delta_2-\delta_1)^2}{\sigma_{\delta,1}^2 + \sigma_{\delta,2}^2}}
\end{equation}

where $\alpha$ and $\delta$ represent the Right Ascension and Declination of a given source, and $\sigma$ is the uncertainty in that coordinate for that source. Assuming cross-matches follow a Rayleigh distribution, we apply an effective cutoff of $r=5.68$, which should provide an acceptable miss rate for counterparts of 1 in $10^7$.

These linkages are assembled and analyzed using the networkx library in Python \citep{SciPyProceedings_11}. This results in a graph generated for each Fermi field which contains our grouped sources as subgraphs. Finally, these subgraphs are categorized into one of the following classes:

\begin{itemize}
    \item Unresolved - contains at most one source at each frequency.
    \item Resolved - contains two or more sources at any frequency
\end{itemize}

\begin{figure*} 
    \centering
    \includegraphics[width=\textwidth]{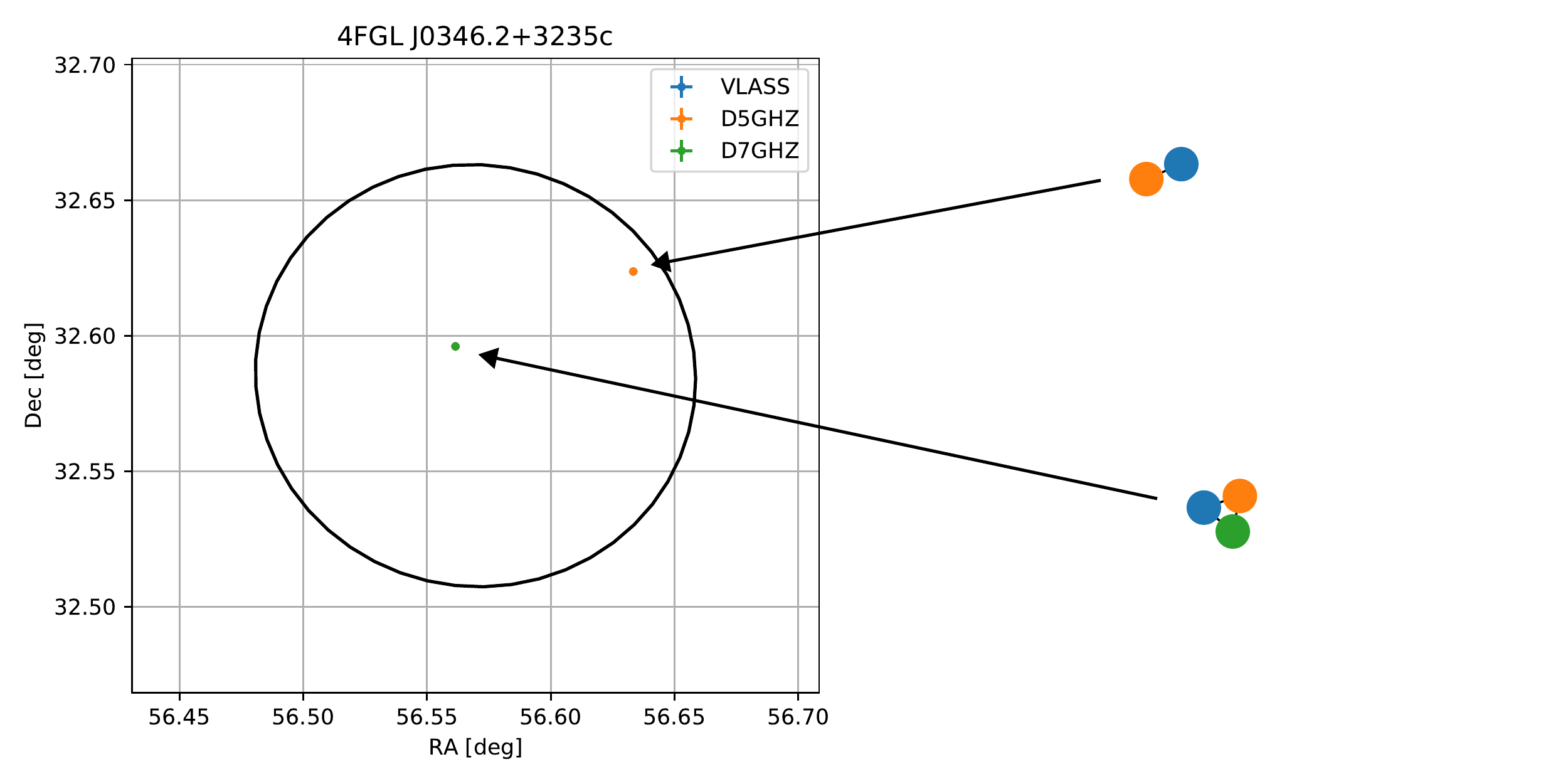}
    \caption{A diagnostic plot generated by our source and spectrum finding pipeline. The left half of the plot shows the 95\% Confidence Positional Uncertainty Ellipse, and all the sources identify inside it from the various catalogs discussed in this paper. The right shows the networked source graphs, created as described in Section \ref{sec: methods}. This format clearly illustrates the two sources multi-frequency nature, as well as hinting to some simplistic notes on the source morphology.}
    \label{fig: exmp_map}
\end{figure*}

\begin{figure*} 
    \centering
    \includegraphics[width=\textwidth]{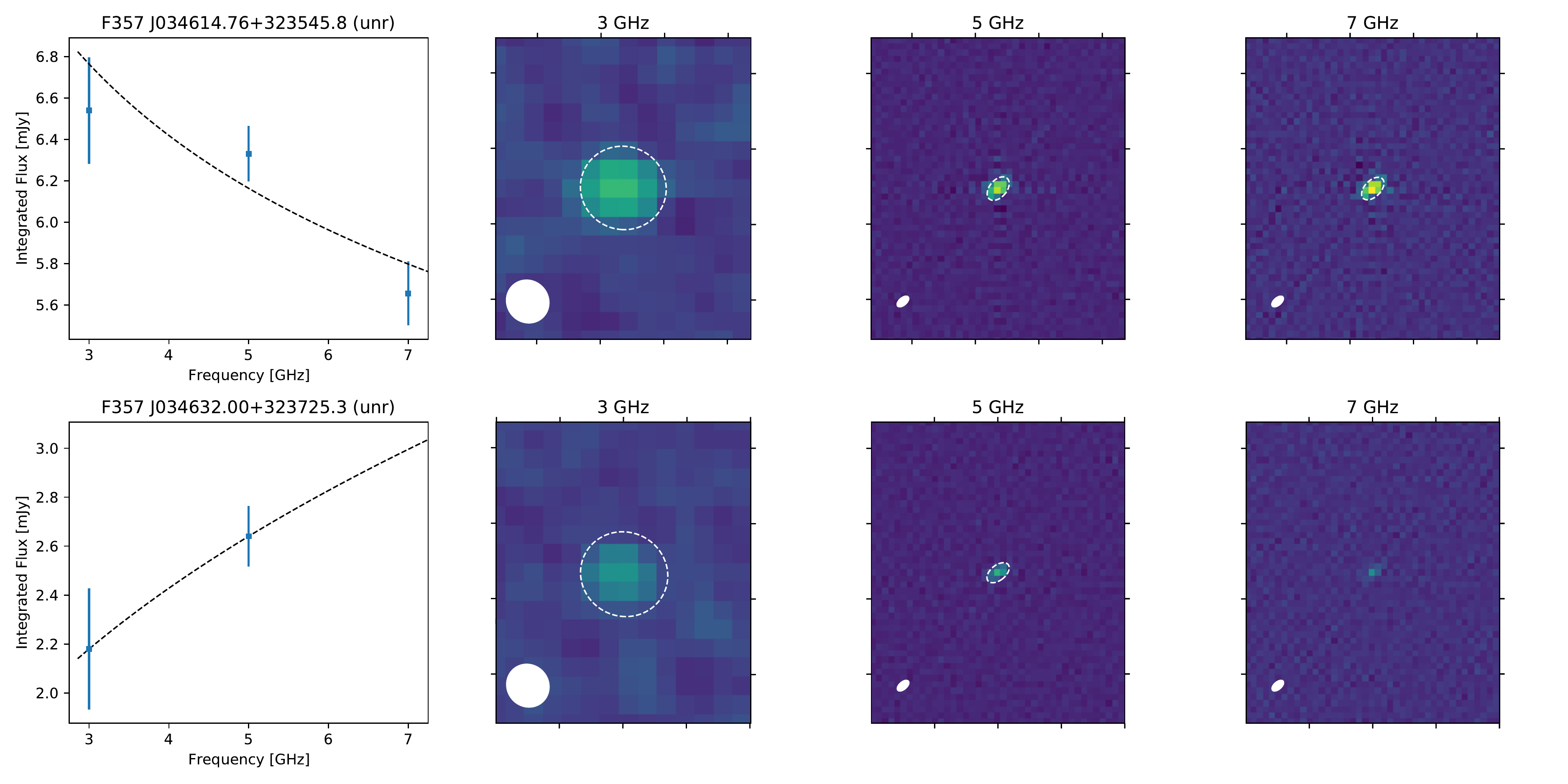}
    \caption{Image cutouts for both of the sources shown in Figure \ref{fig: exmp_map}. Note that the image cutouts are 16x16 arcsec centered around the F357 source location, with the image ticks set to a spacing of 4 arcsec. Beam size is overlaid in the lower-left of each image. We note that diagnostic plots as seen here and in Figure \ref{fig: exmp_map} can be found for every source in our catalog at \url{https://go.nrao.edu/F357}.}
    \label{fig: exmp_ims}
\end{figure*}

This distinction allows us to group sources effectively while accounting for any sort of resolved structure which may appear at any frequency (particularly higher frequencies, which will have an increased resolution). This, combined with other parameters given by PyBDSF, allow us to thoroughly probe the overall structure of sources, without the need to examine them in the actual image. One such diagnostic plot can be seen in Figures \ref{fig: exmp_map} and \ref{fig: exmp_ims}.

For any source in the catalog which contains multi-frequency information (Resolved or Unresolved), we can then identify the source's spectral index (with $S \propto \nu^{+\alpha}$ in this work). For sources with data at two frequencies, a simple two point fit is performed such that

\begin{equation}
    \alpha = \frac{\log{\left( {S_2}/{S_1} \right)}}{\log{\left( {\nu_2}/{\nu_1} \right)}} \text{, and}
\end{equation}

\begin{equation}
    \sigma_\alpha = \frac{ \left( {\sigma_1}/{S_1} \right)^2 + \left( {\sigma_2}/{S_2} \right)^2}{ \log{\left( {\nu_2}/{\nu_1} \right)}^2 } \text{.}
\end{equation}

In the case where data is available at all three frequencies, we perform an uncertainty weighted least-squared minimization via the \texttt{optimize} module in SciPy \citep{2020NatMe..17..261V}. In the case of Resolved sources, spectral index calculations are performed using the sum of integrated flux values at each frequency. It is important to note that this makes the values for Resolved source spectral indices somewhat less reliable, but they are included in our data product for the sake of completeness. We also note that, in the scope of this paper, lower limit values are not included when examining spectral index by either of these methods; all spectral indices may be thought of as determined by well constrained values for flux.

For each source, we also provide the relevant "S-Codes" generated by PyBDSF, combined as a single string with three characters ('S', 'C', or 'M' from PyBDSF, or '-' when no source was identified at that frequency). These values encode information about the Gaussian-fit and it's surroundings, and as such are a valuable way to describe the source morphology at a glance. In example: a TypeCode of 'SM-' represents a source with unresolved emission at 3 GHz, some type of resolved emission at 5 GHz, and a non-detection at 7 GHz. The exact nature of these S-Codes is related to the Gaussian fitting process used during source extraction, and it is recommended those interested view the PyBDSF documentation\footnote{\url{https://www.astron.nl/citt/pybdsf/write_catalog.html}} for full descriptions. Here we provide a cursory summary:

\begin{itemize}
    \item S - An isolated source which can be described by a singular Gaussian.
    \item C - A source which can be described by a single Gaussian, but which exists as part of a larger flux structure (called an Island by PyBDSF).
    \item M - A source best described by multiple Gaussians.
\end{itemize}

As a final step we cross-match our identified sources against several relevant catalogs using TOPCAT's pair matching system \citep{2005ASPC..347...29T}. The utility of these cross matches is discussed more thoroughly for each type of source in Section \ref{sec: results}. In particular, we cross-match against WISE \citep[AllWISE:][]{2010AJ....140.1868W, 2011ApJ...731...53M}, Gaia \citep[DR2:][]{2016A&A...595A...1G, 2018A&A...616A...1G}, ROSAT \citep[2RXS:][]{2016A&A...588A.103B}, Chandra \citep[CSC 2.0:][]{2020AAS...23515405E}, and Swift \citep[2SXPS:][]{2020ApJS..247...54E}. These cross-matches, as well as their separations from our sources, are provided in the final catalog. 

\begin{figure}
    \centering
    \includegraphics[width=\linewidth]{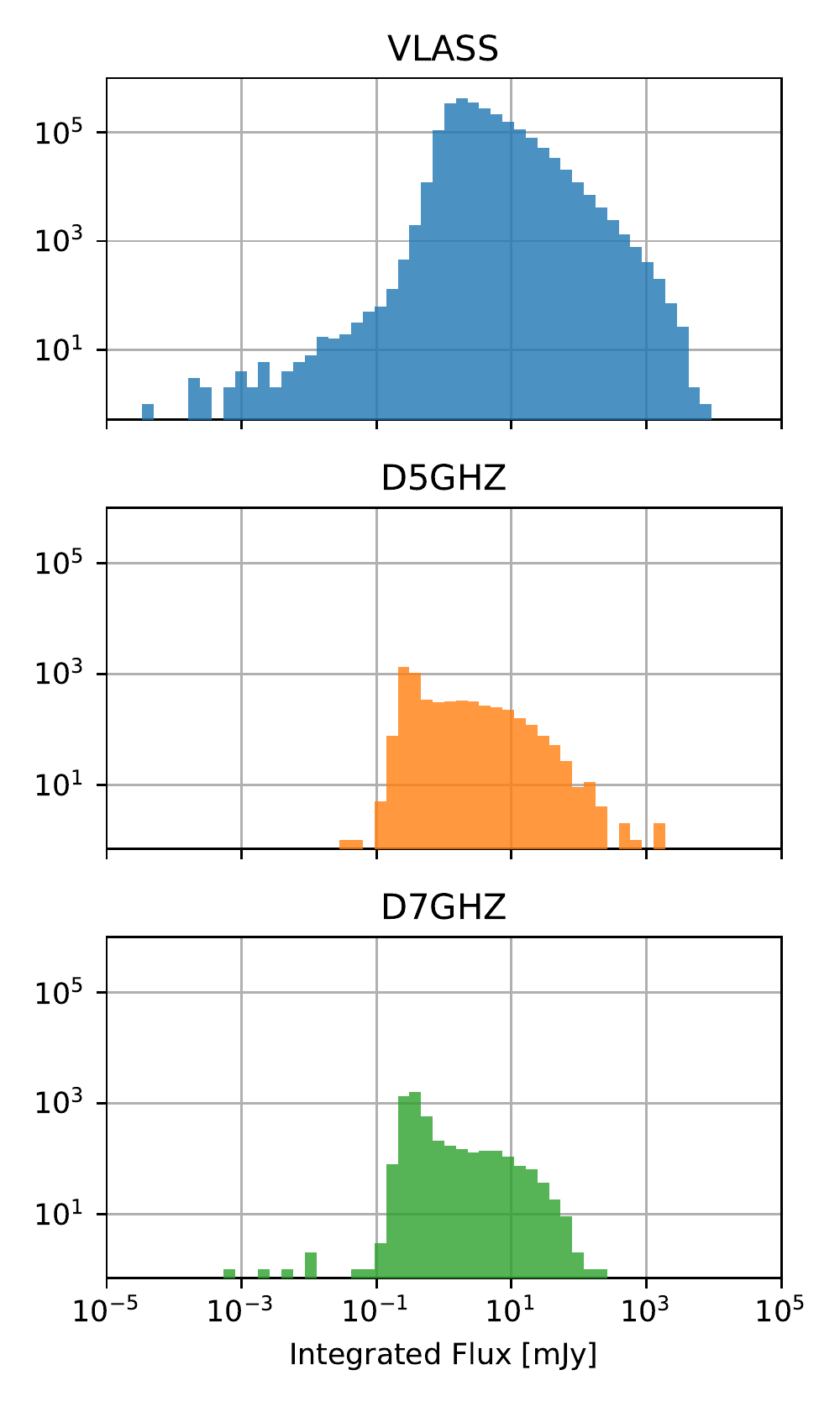}
    \caption{Survey histograms showing the number of sources in logarithmic flux bins. Note that the left end of each histogram represents our completeness limit, the deepest our survey is able to probe. In the case of our 5 and 7 GHz observations, this is where picking out sidelobe sources becomes more difficult for automated systems such as PyBDSF, creating the seeming "bump" in low flux sources in those surveys.}
    \label{fig: logNlogS}
\end{figure}

We also provide here in the form of Figure \ref{fig: logNlogS} a quantitative assessment of our various surveys' completeness and sensitivity. In particular for the 5 and 7 GHz surveys it is important to note that below the completeness limit (around 10 mJy) it becomes difficult for automated systems such as PyBDSF to differentiate real sources from sidelobe sources, causing the artificial bump in the number of sources at the lower limits (around 1 mJy). Sidelobe levels are in general higher for our dedicated C-band observations due to short single snapshot coverage of the Fourier plane, whereas VLASS despite its snapshot nature has significantly better coverage and lower sidelobe levels. It should also be noted that in VLASS "quicklook" images, regions around bright sources are excluded.

\begin{deluxetable*}{cc|cc|cc|cc|c} \label{tbl: sec}
\tablecaption{F357 Secondary Catalog}
\tablehead{\colhead{Field ID} & \colhead{Name} & \multicolumn{2}{c}{VLASS} & \multicolumn{2}{c}{D5GHZ} & \multicolumn{2}{c}{D7GHZ} & \colhead{Coverage\tablenotemark{a}} \\
\colhead{} & \colhead{All} & \colhead{Significant} & \colhead{All} & \colhead{Significant} & \colhead{All} & \colhead{Significant} & \colhead{All}}
\startdata
0 & 4FGL J0000.5+0743 & 2 & 1 & 0 & 0 & 0 & 0 & 0.0000 \\
1 & 4FGL J0003.3+2511 & 1 & 0 & 11 & 0 & 27 & 0 & 1.0000 \\
2 & 4FGL J0003.6+3059 & 1 & 1 & 3 & 0 & 0 & 0 & 1.0000 \\
3 & 4FGL J0004.0+5715 & 2 & 1 & 1 & 1 & 0 & 0 & 1.0000 \\
4 & 4FGL J0005.6+6746c & 12 & 6 & 0 & 0 & 0 & 0 & 0.0400 \\
5 & 4FGL J0006.6+4618 & 2 & 2 & 4 & 2 & 3 & 2 & 1.0000 \\
6 & 4FGL J0008.4+6926 & 2 & 2 & 4 & 2 & 8 & 2 & 1.0000 \\
7 & 4FGL J0008.9+2509 & 19 & 17 & 1 & 1 & 1 & 1 & 0.2302 \\
8 & 4FGL J0009.2+6847 & 1 & 1 & 2 & 0 & 0 & 0 & 0.3987 \\
9 & 4FGL J0009.2+1745 & 2 & 2 & 1 & 1 & 0 & 0 & 0.6052
\enddata
\tablecomments{Here the term `Significant` refers to sources above 1 mJy.}
\tablenotetext{a}{This column refers specifically to the 5 and 7 GHz coverage, as VLASS can be assumed to have approximately complete coverage of the sky above $\delta > -40$ deg.}
\end{deluxetable*}

The resultant data product we call the F357 Catalog (representing \textit{Fermi} and our three radio frequencies). This catalog is the extension of nearly a decade of work surveying Northern sky unassociated gamma-ray sources, and as such provides perhaps the most in depth radio analysis of these fields as a whole. To qualify this, we also provide the F357 Secondary Catalog, which provides metadata on the relative coverage of each field, including the number of sources (and significant sources, above 1 mJy) at each frequency, as well as the ellipse coverage fraction for our dedicated observations. The first 10 lines of said catalog can be seen in Table \ref{tbl: sec}, and the full table, the F357 catalog, the VLASS quicklook catalog, and source image cutouts can be obtained at \url{https://go.nrao.edu/F357}.

\section{Results} \label{sec: results}

Previous iterations of this work have made significant contributions in the form of AGN associations. Initial observations at 5 and 7 GHz were used to identify potential candidates of interest, which were then targeted using VLBI observations so as to resolve structure scales indicative of AGN. Once a target was confirmed to be an AGN, it could then be given an association likelihood based on its proximity and brightness.

Owing to our VLASS quicklook catalog, we were able to include robust spectral data. Such data were not available before for $\gamma$-ray association of weak sources. Using this, we are able to analyze the spectra of a significantly larger number of sources than the previous two-point data would have ever allowed for. 

\begin{figure*} \label{fig: cmalpha}
    \centering
    \includegraphics[width=\textwidth]{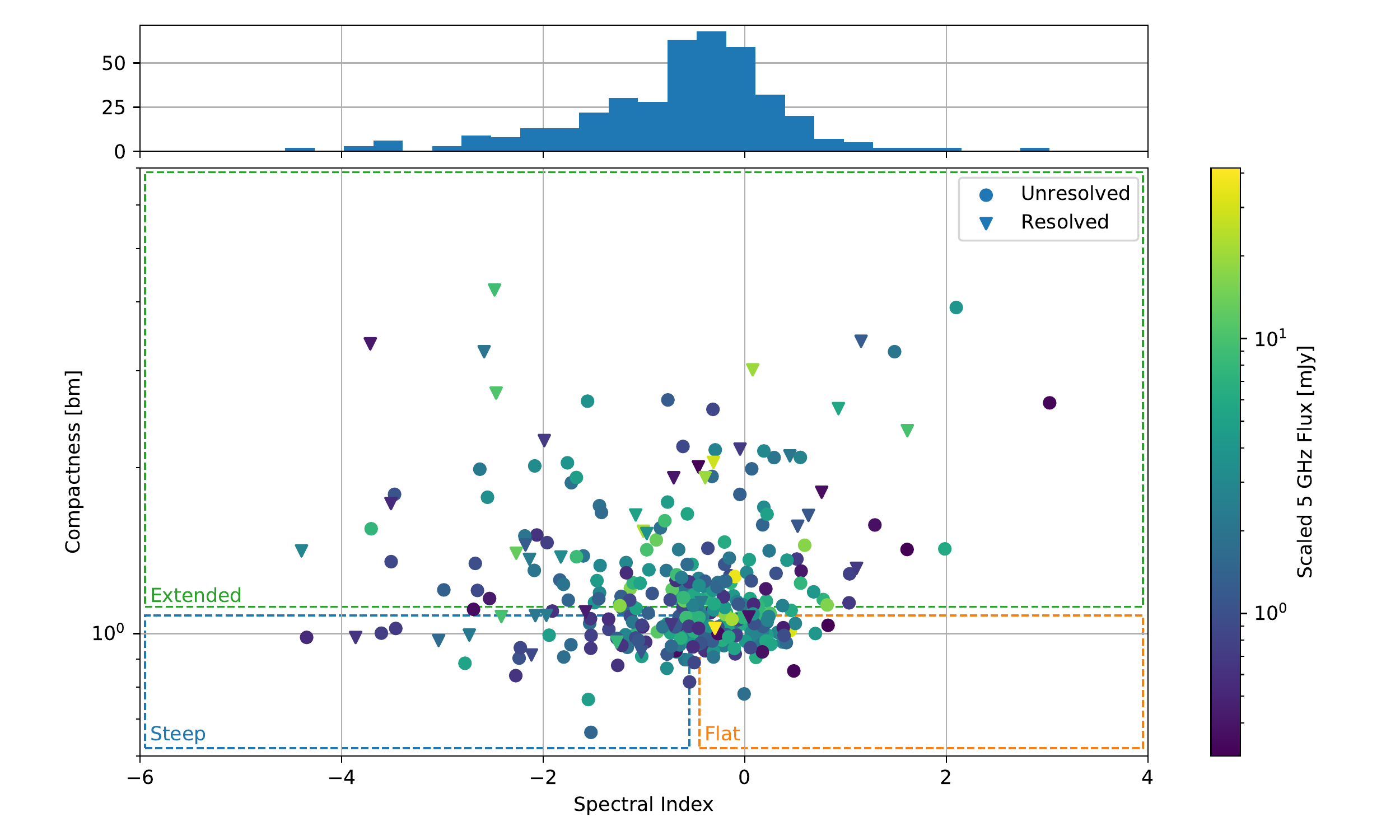}
    \caption{Properties of sources with spectral index information in both the Resolved (triangular points) and Unresolved (circular points) F357 Catalogs. The dashed squares overlayed on the plot represent the various source classes discussed in Section \ref{sec: results}, and match the color of sources seen in Figure \ref{fig: skymap}.}
    \label{fig:cmalpha}
\end{figure*}

Furthermore, via both the PyBDSF source morphology code (S\_Code) and the ratio of integrated to peak flux density (hereafter referred to as the Compactness Measure, or CM) we have a means to probe the structure/size of any particular source. With this in mind, in the first four subsections to follow, we split our discussion around three distinct regions in the space of spectral index and compactness measure (as can be seen in Figure \ref{fig: cmalpha}), along with some particularly interesting exceptions to these regions. These correspond to the designations given in the Class column of our F357 catalog (see Table \ref{tbl: f357cols}).

\subsection{Steep Spectrum Sources} \label{sec: stp}

This first class of sources has spectral index $\alpha<-0.5$, and compactness measure $\text{CM}<1.1$, representing compact steep-spectrum sources. A similar classification was used in \citet{2018MNRAS.475..942F} to identify pulsar candidates, as the pulsar spectral distribution peaks at approximately $\alpha=-1.4$ \citep{2013MNRAS.431.1352B}. For comparison, the sample that is flux limited at a 2 mJy level at 150 MHz has a median spectral index between 150 MHz and 1.4 GHz of $\bar{\alpha}=-0.78$ \citep{2017A&A...598A..78I}. Among our sample of spectral indices, we do not note a strong separation between these two populations, and as such we adopt the median spectral index $\tilde{\alpha}=-0.5$ (rounding from -0.497) as our steep-spectrum cutoff.

This method represents a powerful tool in identifying sources which have been previously hidden from pulsar-timing surveys. There are a variety of means by which a pulsar's timing signal may be lost in transit or difficult to identify, while the radio spectrum is preserved; namely obscuration by the intervening medium (see Section \ref{sec: empty} for a more thorough discussion of these types of sources) and pulsars in binary systems. While the overall binary fraction for young pulsars is somewhat low \citep{2021MNRAS.501.1116A}, recycled millisecond pulsars by definition exist in binary systems, and this can complicate pulsar timing due to necessitating more advanced orbital models to extract the periodic signal. Such systems are relatively well represented in 4FGL associations and identifications \citep[e.g.][]{2020ApJ...892...21S, 2020ApJ...904...49M}.

It is important to note that the classification of a source as compact and steep spectrum does not necessarily identify it as an obvious pulsar; other sources can sometimes manifest steep-spectra. A primary contaminate noted in \citet{2018MNRAS.475..942F} and \citet{2008A&ARv..15...67M} are luminous high redshift radio galaxies (HzRGs). These objects, while interesting, are not likely to be $\gamma$-ray sources. Differentiation between pulsars and HzRGs requires follow-up at near-milliarcsecond resolution, which can identify kpc-scale structure in the HzRGs, whereas pulsars will still remain completely unresolved at these scales.

With the above concerns noted, these sources can be identified by the `STP` label in the `Class` column in the F357 catalogs. We note that for consideration of pulsar-candidates, one need only reference the F357 Unresolved catalog, as we should not expect pulsars to exhibit structure on the scale of our resolution. Table \ref{tbl: pulsars} shows the full list of 39 compact unresolved steep-spectrum sources.

We also note that it is possible to infer a source's spectral shape to some degree. We perform this operation for sources which identify a VLASS source, but do not find an associated source at 5 or 7 GHz in fields with coverage of at least 90\%. In these cases we can calculate an upper limit on the value of the spectral index $\alpha$ using the completeness limit of our survey (10 mJy) as an approximate upper limit on the flux of the source there. From these values, we then identify those sources with a spectral index $\alpha<-0.5$, and identify these as "Inferred Steep Spectrum", applying the `IFR` label in the `Class` column. These sources are also included in Table \ref{tbl: pulsars}.

These sources are of interest for similar reasons to the rest of the `STP` classification; pulsars (or ostensibly other steep spectrum emission) may be scatter by various mechanisms, which we may expect to have a more pronounced effect at higher frequencies. A pulsar viewed through significant scattering material (here the galactic interstellar medium) may well show up in the lower frequency (VLASS) observations, but be hidden from our higher-frequency 5 and 7 GHz observations.

\begin{deluxetable*}{ccccccc}
\tablecaption{Steep Spectrum Sources}
\tablehead{\colhead{Name} & \colhead{l} & \colhead{b} & \colhead{Total Flux Density (5 GHz)} & \colhead{Spectral Index $\alpha$} & \colhead{TypeCode} & \colhead{CrossMatches}\\ \colhead{ } & \colhead{deg} & \colhead{deg} & \colhead{mJy} & \colhead{ } & \colhead{ } & \colhead{ }}
\startdata
F357 J182817.03-325520.8 & 0.9087 & -9.93220 & $3.129 \pm 0.225$ & $-1.552 \pm 0.403$ & SSS & AllWISE \\
F357 J182013.04-292751.2 & 3.2486 & -6.83770 & $0.777 \pm 0.131$ & $-1.021 \pm 0.324$ & SS- &  \\
F357 J180939.47-241324.8 & 6.7706 & -2.28947 & $1.812 \pm 0.190$ & $-1.796 \pm 0.178$ & SS- & CSCR2 \\
F357 J180947.88-241049.1 & 6.8240 & -2.29662 & $0.638 \pm 0.099$ & $-2.272 \pm 0.307$ & SS- &  \\
F357 J153137.47+040739.6 & 9.0530 & 45.42713 & $0.716 \pm 0.118$ & $-2.239 \pm 0.516$ & SSS & AllWISE \\
F357 J181936.89-203631.6 & 11.0486 & -2.59069 & $0.526 \pm 0.095$ & $-11.270 \pm 0.253$ & SS- &  \\
F357 J173857.88-105638.2 & 14.6551 & 10.67087 & $< 10.000$ & $<-0.964$ & S-- &  \\
F357 J173857.89-105638.2 & 14.6551 & 10.67083 & $5.370 \pm 1.066$ & $-2.776 \pm 0.353$ & -SS &  \\
F357 J181022.52-141901.5 & 15.5301 & 2.34527 & $< 10.000$ & $<-0.523$ & S-- &  \\
F357 J161210.62+040937.8 & 16.4520 & 36.97393 & $0.869 \pm 0.239$ & $-2.227 \pm 0.710$ & -SS & AllWISE
\enddata
\tablecomments{Here we provide the 5 GHz Total Flux Density as it is available for all STP class sources.  For the IFR class sources we report the upper limit as determined from our survey characteristics.  The 3 GHz total flux density for such sources can be calculated from the given limits, or found in the full F357 catalog.}
\end{deluxetable*}
 \label{tbl: pulsars}

\subsection{Flat Spectrum Sources} \label{sec: flt}

This class is composed of sources having $-0.5<\alpha$ and CM $<1.1$. These may be considered compact, flat-spectrum sources. Of these, the source type of primary interest will be active galactic nuclei (AGN), which represent a majority of associations throughout the history of the FGL catalogs. 

These sources can be found in the F357 catalogs under the `FLT` label in the `Class` column. We note again that this classification does not on its own identify a source as an AGN; e.g. on the fainter end of this population star forming galaxies will make a significant contribution, which are not expected to be a significant source of $\gamma$-ray emission. The utility of this classification is that it provides a narrowed list of targets, instead of requiring that our entire F357 catalog be surveyed for AGN-targeting observations. If such observations (typically performed using the VLBA) are able to identify parsec-scale emission, then we can adequately begin to evaluate the source's likelihood of being associated.

\subsection{Peculiar Sources} \label{sec: pcl}

Another class we label as Peculiar (PCL). These are sources where the spectrum seems discontinuous, either because the source may be variable (keeping in mind that the 3 GHz observations did not occur simultaneously with the 5/7 GHz observations) or because the spectrum exhibits a turnover. With only three points of spectral/temporal data, we do not attempt to distinguish between these cases, opting instead to simply label such sources. A total of 65 may be found in the F357 catalog.

We identify these Peculiar sources by performing three separate two-point fits for every source with flux values at all three frequencies. For each two-point fit, we note the distance from the non-included point to the fit line, normalizing it by the uncertainty of the flux value. This produces three error-scaled offsets per source, and said source is marked as Peculiar if the average of these offsets $\bar{d}>5$. This criterion roughly picks out sources which have points that deviate from a standard power-law by factors of $5\sigma$ or more. 

\subsection{Extended Sources} \label{sec: ext}

The next class of sources are those having CM $>1.1$. These may be considered extended sources. In this case "resolved" means that more than one source was identified at one or more frequencies by our graph analysis method (see Section \ref{sec: methods}), whereas "extended" here simply refers to sources which meet the above compactness measure criterion. 

As spectral index can be somewhat less useful as a diagnostic for extended sources, our classification scheme applied in the catalog makes no distinction between extended sources with steep-spectra and those with inverted-spectra, for example. The spectral index and its uncertainty are still included in the catalog for the reader's use, keeping in mind the above warnings. Another column in the catalog of note for these sources is `TypeCode`, which provides all the available TypeCodes generated by PyBDSF's source finding (e.g. `SS-` means PyBDSF classified the 3 and 5 GHz sources as singular, but no 7 GHz source was detected). This qualitatively describes the morphology of the source at each frequency. The catalog then offers three ways (compactness measure, resolved/unresolved, and TypeCode) to probe these extended sources.

These sources are still interesting from the perspective of providing associations. While a majority of associations and identifications come from pulsars and AGN, which will appear compact at our observed frequencies, the 4FGL-DR2 catalog and its predecessors have always contained some number of supernova remnants (SNR) and other such sources which will appear as extended when observed in the radio regime. With this in mind we did perform cross-match with the \citet{2019JApA...40...36G} SNR catalog, but identified no matches between our extended sources and known SNR.

\subsection{Empty Fields} \label{sec: empty}

\begin{figure*} 
    \centering
    \includegraphics[width=\textwidth]{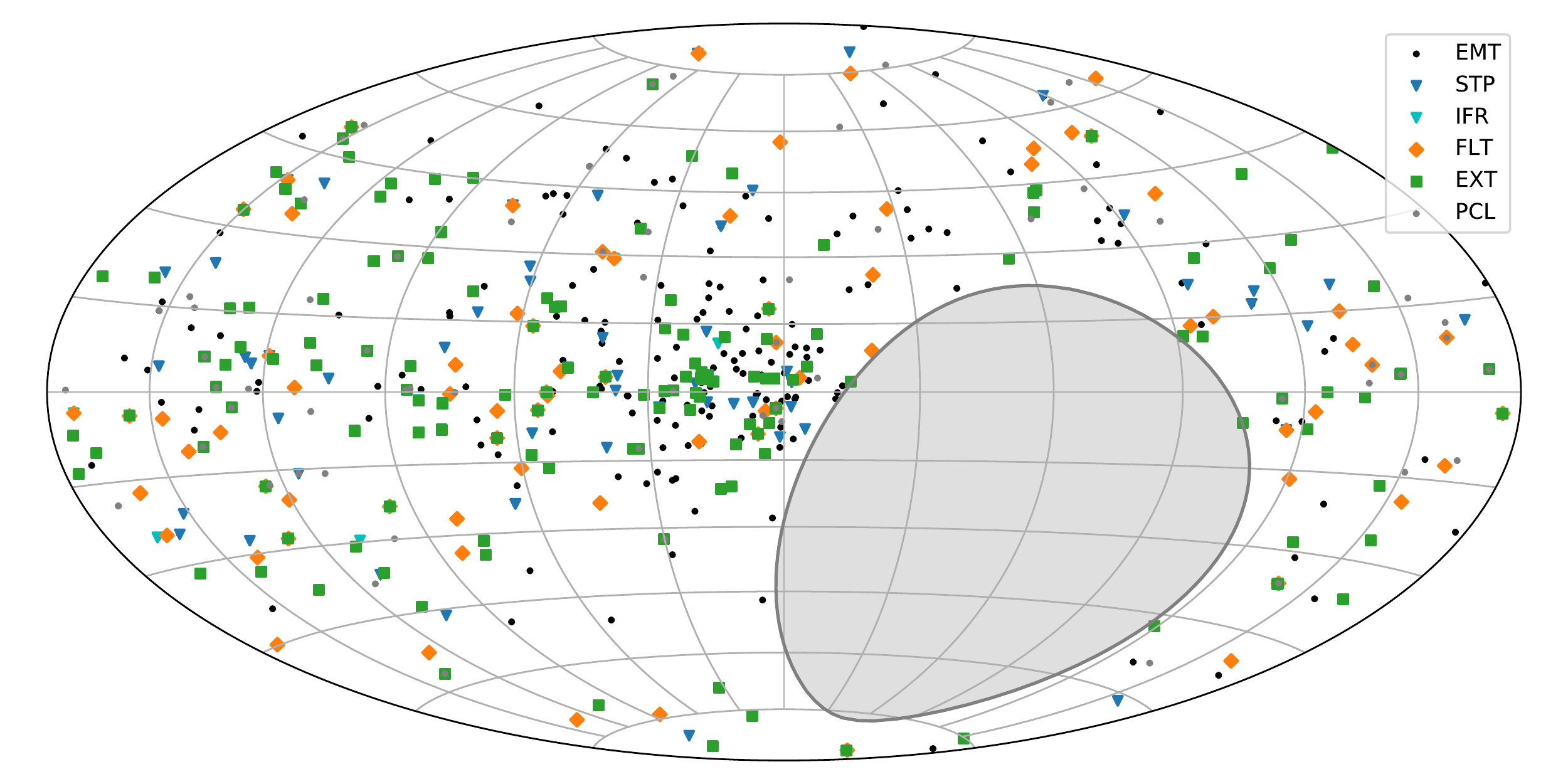}
    \caption{A skymap in galactic coordinates highlighting the distribution of various source classes as discussed in \ref{sec: results}, as well as our coverage area (with the grey region representing the Southern sky obstructed by the Earth). All Empty Fields are shown, along with any source which could be classified via its spectral index and compactness, as denoted in the Legend.}
    \label{fig: skymap}
    \label{fig:sky}
\end{figure*}

Finally we come to unassociated fields where we identify no sources in VLASS or our dedicated observations. Such Empty Fields were first noted in our 3FGL analysis \citep{2017ApJ...838..139S}, and we adopt similar criteria here. A field is classified as empty if it contains no significant sources (having peak flux above 1 mJy/beam) at 5 or 7 GHz, while also being 90\% or more covered by our dedicated observation pointings. As VLASS has nearly complete northern-sky coverage, with the exception of cutouts around selected bright sources, we can assume it has complete coverage of our fields, simplifying our coverage criteria somewhat. With these criteria in mind, we identify 220 Empty Fields, the positions of which can be seen in Figure \ref{fig: skymap}. 

The sky distribution of these sources is of some note. As with previous results, we find that a large fraction of these sources (here 40\%) lie within 10 degrees of the Galactic plane. This is in good agreement with our previous values for 3FGL, and seems to suggest that these sources are primarily related to Galactic counterparts. From the steep-spectrum radio emission (inferred from our lack of detection) and the aforementioned galactic distribution, we consider the possibility that these fields are likely Galactic-plane pulsars, and thus ideal targets for pulsar searches. We note here that these Empty Fields show no real distinction from other gamma-ray sources in 4FGL in terms of their gamma-ray properties, further suggesting they are typical sources that simply require more detailed radio analysis.

Efforts have been made to target these Empty Fields with deeper observations. In particular we have an ongoing effort targeting 3FGL Empty Fields with high pulsar likelihoods in \citet{2016ApJ...820....8S}. These observations were performed with the VLA at 1-2 GHz (L-band), and probe the fields significantly deeper than time allows with our full-catalog surveys. A full analysis of these fields is ongoing and in preparation, but our pilot observations (project 17B-384) has provided a significant number of new association candidates. These include some quite exotic sources, such as a "cannonball" pulsar \citep{2019ApJ...876L..17S} and a somewhat more mysterious radio source which had pulses detected by the realfast system \citep{2018ApJS..236....8L} running concurrent with our observations. The exact nature of the latter source is still unknown, and further observations are ongoing to learn more about this potential association candidate. So far, at least 10 new pulsars have been found independently in Empty Fields following their discovery by \citet{2017ApJ...838..139S}.

\section{Discussion \& Summary} \label{sec: discussion}

As predicted in our 3FGL analysis \citep{2017ApJ...838..139S}, continued efforts at radio frequencies have been able to push the completeness limit on $\gamma$-ray radio counterpart searches down to about 10 mJy. This resulted in a reduction of unassociated sources in 4FGL to a fraction of about 26\%. With 4FGL-DR2 probing deeper into the $\gamma$-ray sky, new fainter unassociated sources have been picked up. Together with the catalog presented here, we are able to present a significant number of new radio sources existing inside localization ellipses of unassociated 4FGL-DR2 sources and, leveraging the extensive coverage of VLASS, we can provide more reliable spectral information on many of these sources, which can be used to further aid in efforts toward association. Considering the relatively large number of sources presented in our newest catalog, this additional spectral information allows us to classify sources based on their spectrum, and when combined with image-plane properties such as the measure of compactness, we gain the ability to specifically target certain types of sources during specific follow-up programs focusing on different $\gamma$-ray systems (e.g. AGN, pulsars, etc...). 

Such specific targeting becomes more relevant as we continue to push the limits of radio associations to these objects. While there are ongoing efforts to classify such sources based on their $\gamma$-ray properties \citep[see][]{2016ApJ...820....8S,2021JHEAp..29...40C}, such methods typically become more difficulty when the overall $\gamma$-ray flux of the source is low, as is the case with most of the new unassociated sources added to the FGL catalog in recent years. As \textit{Fermi} continues to collect data, newly identified unassociated sources and certain persistently unassociated sources such as our Empty Fields, will require yet deeper and more extensive radio searches to extract association candidates. As our completeness limit is pushed down, the total number of sources found inside these fields by these searches will increase, and some method for narrowing down targets for follow-up becomes necessary to keep false associations at a minimum and to allow for efficient follow-up observations, e.g. through VLBI for AGN associations. 

Our catalog provides a large list of potential sources for association follow-up. Beyond that, it also provides a means by which to actively target specific desired source types using follow-up that is more relevant for that particular source type. For example, compact steep spectrum sources or Empty Fields may be targeted using single-dish instruments or low-frequency interferometric arrays, such as our current follow-up project using the Long Wavelength Array \citep[LWA;][]{2012JAI.....150004T} to target a selection of the steep-spectrum targets presented in Table \ref{tbl: pulsars}. We also target compact flat-spectrum sources with the VLBA to search for milliarcsecond-scale structure hinting at an AGN association. Different source progenitors are best targeted using somewhat different methods, and the work presented here allows for those selections to be made for a sizable fraction of our 7432 presented sources.

Furthermore, even a lack of sources provides a unique sort of classification. We mentioned in Section \ref{sec: empty} our specific efforts to target Empty Fields. Obscured $\gamma$-ray pulsars represent perhaps the most likely progenitor for a large number of these particular fields, and as such we've targeted them with deep VLA observations at 1.4 GHz (L-band). Early results from these observations have been quite successful at identifying likely associations and even identifications in some sources, and reduction for the majority of these fields is still ongoing. We note that these Empty Field observations were performed using targets from our 3FGL analysis, and between these fields and new ones generated by the analysis presented in the work, we expect to continue to produce a significant number of association targets from this particular class of field.

Beyond the above mentioned follow-up, we also intend to expand upon the spectral analysis detailed in this work by leveraging yet more catalogs available at a variety of frequencies, leveraging combination with e.g. RACS \citep{2020PASA...37...48M} or TGSS \citep{2017A&A...598A..78I}. By incorporating fluxes and limits from these and other surveys we gain increased spectral resolution, particularly towards sources that peak at lower frequencies, such as our steep-spectrum candidates. With an increased number of data points, we can also make use of more sophisticated spectrum fitting systems, such as a full Markov-Chain Monte Carlo (MCMC) treatment. This work is ongoing and is built from the generic source matching and spectrum fitting system created for this work.

In summary, here we provide the F357 catalog, containing radio sources identified inside localization ellipses for unassociated \textit{Fermi} sources. This catalog includes new sources only recently available to us via VLASS, and as such we also provide our separate VLASS quicklook catalog (see Appendix \ref{apn: vlass}) which may be of general use to the community. Leveraging VLASS, along with our multi-frequency dedicated observations, we are able to provide spectral information and some degree of source classification for a significant number of sources in our F357 catalog, allowing for more efficient follow-up toward continuing efforts to associate and identify a persistently mysterious fraction of the $\gamma$-ray sky.

\section{Acknowledgements}

We thank Dale Frail for useful discussions in the context of this paper. SB, FKS, and GBT, acknowledge support by the NASA Fermi Guest Investigator program, grants 80NSSC19K1508, NNH17ZDA001N, NNX15AU85G, NNX14AQ87G, and NNX12A075G. The National Radio Astronomy Observatory is a facility of the National Science Foundation operated under cooperative agreement by Associated Universities, Inc.

This publication makes use of data products from the Wide-field Infrared Survey Explorer, which is a joint project of the University of California, Los Angeles, and the Jet Propulsion Laboratory/California Institute of Technology, and NEOWISE, which is a project of the Jet Propulsion Laboratory/California Institute of Technology. WISE and NEOWISE are funded by the National Aeronautics and Space Administration. This work has made use of data from the European Space Agency (ESA) mission {\it Gaia} (\url{https://www.cosmos.esa.int/gaia}), processed by the {\it Gaia} Data Processing and Analysis Consortium (DPAC, \url{https://www.cosmos.esa.int/web/gaia/dpac/consortium}). Funding for the DPAC has been provided by national institutions, in particular the institutions participating in the {\it Gaia} Multilateral Agreement. This research has made use of data obtained from the Chandra Source Catalog, provided by the Chandra X-ray Center (CXC) as part of the Chandra Data Archive. We acknowledge the use of public data from the Swift data archive.

This research has made use of NASA’s Astrophysics Data System and has made use of the NASA/IPAC Extragalactic Database (NED) which is operated by the Jet Propulsion Laboratory, California Institute of Technology, under contract with the National Aeronautics and Space Administration. This research has made use of data, software and/or web tools obtained from NASA’s High Energy Astrophysics Science Archive Research Center (HEASARC), a service of Goddard Space Flight Center and the Smithsonian Astrophysical Observatory, of the SIMBAD database, operated at CDS, Strasbourg, France.

\newcommand{\furl}[1]{\footnote{\url{#1}}}

\software{AIPS\furl{www.aips.nrao.edu} \citep{2003ASSL..285..109G},
          Astropy\furl{www.astropy.org} \citep{2013A&A...558A..33A},
          CASA\furl{www.casa.nrao.edu} \citep{2007ASPC..376..127M},
          matplotlib\furl{www.matplotlib.org} \citep{2005ASPC..347...91B},
          networkx\furl{www.networkx.org} \citep{SciPyProceedings_11},
          Numpy\furl{www.numpy.org} \citep{2011CSE....13b..22V},
          Scipy\furl{www.scipy.org} \citep{2020NatMe..17..261V},
          TOPCAT\furl{www.star.bris.ac.uk/~mbt/topcat} \citep{2005ASPC..347...29T}}

\appendix
\section{The VLASS Quicklook Catalog} \label{apn: vlass}

Here we present a description of VLASS observations and data used to produce our Quicklook Catalog. A full description of the VLASS project can be found in \citet{2020PASP..132c5001L}, but in summary: the entire northern sky (above $\delta = 40$ deg) is surveyed at 3 GHz during B configuration. This process is repeated for three separate epochs which can be used for future transient searches. As a whole, VLASS provides certainly the deepest and highest resolution wide-area survey of the northern sky at these frequencies to date.

The main data products from the reduction pipeline are intended to be fully descriptive VLA data, including spectral and polarization information. While the reduction process is being refined, the pipeline currently outputs simple total-intensity maps, which are called "quicklook" images. These are likely to be fairly true when eventually compared with the final data product, with the only particular exception being a marginally higher RMS noise. As of writing, observations for Epoch 1 have been completed, and those for Epoch 2 are currently ongoing. Our catalog, provided alongside this paper, identifies all sources in the VLASS Epoch 1 quicklook images.

These identifications are performed using PyBDSF, scripted via Python to process the data in parallel on the NRAO's New Mexico Post Processing Cluster (nmpost). For the reduction of VLASS images, as well as our dedicated observations, we use the following parameters in the \texttt{process\_image} method:

\begin{verbatim}
    rms_box = (60,20)
    adaptive_rms_box = True
    rms_box_bright = (20,6)
    thresh_isl = 3
    thresh_pix = 5
\end{verbatim}

We also note that it is important to set \texttt{catalog\_type='srl'} in the \texttt{write\_catalog} method to ensure that PyBDSF outputs a list of sources, instead of the list of Gaussians which make them up. This source list is more useful for describing astrophysical sources than the Gaussian list PyBDSF outputs by default as of Version 1.9.2. This then produces a source list for each of the quicklook images, and so we then group these into a singular catalog containing roughly 2.2 million sources. As there is some overlap between images in the survey, we perform internal cross-matching to identify sources which appear twice because of these overlapped regions. Both sources are kept for the sake of retaining all possible information, but we apply a flag in the `Flag` column to both. A `1` flag denotes the source in the overlapping pair with the lower peak flux error, while a `2` flag is given to the higher peak flux error. There are also more rarely circumstances where overlaps occur between different observation sessions, in which case the time separation between the sources could prove interesting, and thus both are given the `1` flag. Again, all sources are included, as well as the metadata which one would need to apply their own masking or flagging criteria if that were required.

\begin{figure} \label{fig: offsets}
    \centering
    \includegraphics[width=0.9\textwidth]{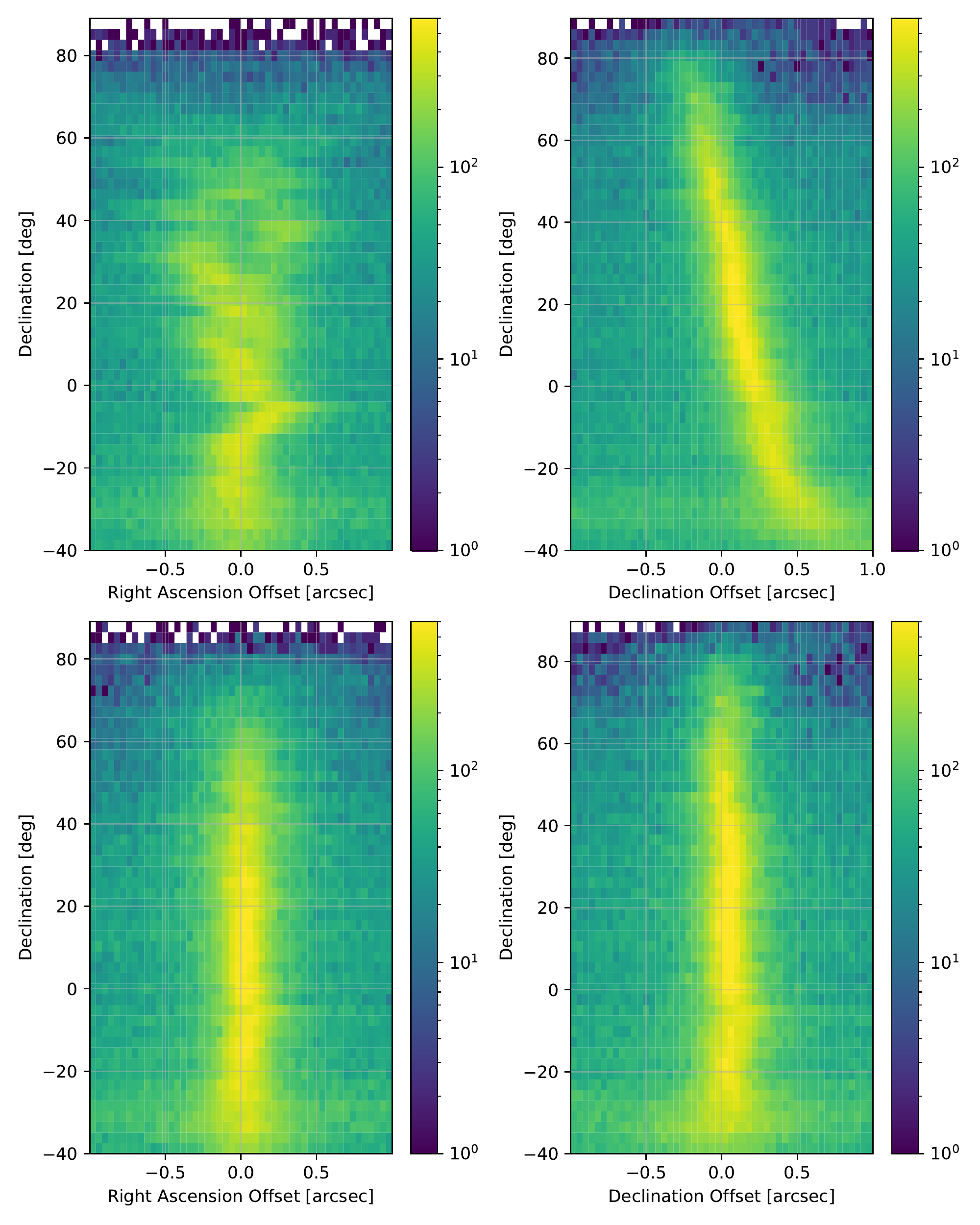}
    \caption{Offsets in right ascension (scaled by $\cos\delta$) and declination as a function of declination. In all cases, offsets are calculated by subtracting the relevant coordinate of the GAIA cross match from the VLASS coordinate (e.g. $\delta_{VLASS} - \delta_{GAIA}$). (Top) The uncorrected offsets. (Bottom) Offsets with corrections in hour angle and declination applied. The remaining offsets present at $\delta\approx-10$ deg are likely due to the equatorial geostationary satellite belt.}
\end{figure}

The final step we apply is a correction of coordinates. It was noted in VLASS Memo 13\footnote{The entirety of the VLASS Project Memo Series can be found at \url{https://go.nrao.edu/vlass-memos}} that during comparison with Gaia sources that there appeared to be a systematic offset in the positions of the VLASS sources. We then performed cross-matching against Gaia DR2 sources, generating a list of offsets for sources with cross-matches within 3 arcseconds. Analyzing these offsets, we found the Declination offset to be a function of Declination, as this was easily corrected with a simple fit. The offsets in Right Ascension at first seem somewhat more chaotic, but reveal a much simpler relation as a function of hour angle. Fitting this as well, we apply our corrections, and propagate the relevant uncertainties. The effects of this correction can be seen plainly in Figure \ref{fig: offsets}. As these offsets seem to be functions of pointing position on the sky, it is currently believed that they will be remedied by the more thorough reduction process for the final VLASS data that will take these projection effects into account.

\begin{deluxetable}{lccl}
\caption{VLASS Columns}
\tablehead{\colhead{name} & \colhead{format} & \colhead{unit} & \colhead{description}}
\startdata
Name & 24s &  & Generated source name \\
RA & 3.4f & deg & Right Ascension of source (J2000 \\
RAErr & 1.2e & deg & Error in Right Ascension \\
Dec & 2.5f & deg & Declination of source (J2000) \\
DecErr & 1.2e & deg & Error in Declination \\
IntFlux & 1.2e & Jy & Integrated flux of source \\
IntFluxErr & 1.2e & Jy & Error in integrated flux \\
PeakFlux & 1.2e & Jy / beam & Peak flux of VLASS source \\
PeakFluxErr & 1.2e & Jy / beam & Error in peak flux \\
TypeCode & 1s &  & Source morphology code from PyBDSF \\
Pointing & 44s &  & Pointing/image information for source \\
ObsDate & 26s &  & Datetime when source was observered \\
Flag & 1d &  & Flags (0=Alone,1=Useful,2=Redundant)
\enddata
\tablecomments{Here the format column refers specifically to the Python string formatting code used to generate the final output table. Boolean columns have no specified formatting.}
\end{deluxetable}
 \label{tbl: vlass}

This final catalog, with our positional corrections applied, is available for general use as a part of this work. Table \ref{tbl: vlass} also provides a description of the columns in our catalog. We note that an alternative VLA sky survey quicklook catalog has been made available while our work was in preparation \citep{2020RNAAS...4..175G}. The major difference between this catalog and the one we present here is that it did not apply astrometric source position corrections as described above. We similarly did not correct for the apparent slight underestimation of flux densities of the order of $\sim$10\%.

\end{document}